\documentclass[a4paper,12pt,english,german]{article}

\usepackage{mathtools}
\usepackage{graphicx}

\begin{document}

\centerline{\bf  Beyond Complementarity}
\bigskip

\centerline{ R. E. Kastner\footnote{Foundations of Physics Group, University of Maryland, College
Park, USA}\footnote{rkastner@umd.edu}}

 \centerline{6 March 2016}

\bigskip

{\small ABSTRACT. It is argued that Niels Bohr ultimately arrived at positivistic and antirealist-flavored statements because of weaknesses in his initial objective of accounting for measurement in physical terms. Bohr's investigative approach faced a dilemma, the choices being (i) conceptual inconsistency or (ii) taking the classical realm as primitive. In either case, Bohr's `Complementarity' does not adequately explain or account for the emergence of a macroscopic, classical domain from a microscopic domain described by quantum mechanics. A diagnosis of the basic problem is offered, and an alternative way forward is indicated.}

\bigskip

\noindent {\bf 1. Introduction. }\\

In this volume\footnote{\textit{Quantum Structural Studies}, eds. R.E. Kastner, J. Jekni\'c-Dugi\'c, G. Jaroszkiewicz, World Scientific Publishers, forthcoming.}, Bai and Stachel [1] offer a rebuttal of arguments by Beller and Fine [2] that Bohr's
philosophy of quantum mechanics was positivist. That discussion addresses Bohr's reply [3] to the Einstein, Podolsky and Rosen (`EPR') paper [4] . The purpose of the present paper is not to enter into the specific debate concerning whether Bohr's basic approach was positivist or not (although this author tends to agree with Bai and Stachel that Bohr's interpretive intentions were not antirealist.) Rather, the intent is to argue that Bohr inevitably lapsed into antirealist-flavored statements about quantum systems because his notion of ``Complementarity'' cannot consistently account for the emergence of classicality from the quantum level. It is argued that ultimately this problem arises from Bohr's implicit assumption that all quantum evolution is unitary; i.e., that there is no real, physical non-unitary collapse.

It should be noted that Bohr's ideas changed and evolved over several decades and this paper does not
attempt to trace the intricate development of this evolution. Rather, attention is focused on Bohr's initial reply to EPR and on certain methodological and metaphysical constraints that, it is argued, led inexorably to a final antirealist position toward quantum level, as evidenced in his famous statement ``There is no quantum world. There is only an abstract quantum mechanical description.'' [7] While a reader might disagree with whether Bohr was instrumentalist or antirealist at any particular stage of the development of his thought, the point of this paper is to argue that the end result of Bohr's investigations into the problem was a form of antirealism about the quantum level that is not in fact forced on us but arises from certain unacknowledged metaphysical, theoretical and methodological assumptions which acted as unnecessary constraints on his interpretive investigation (and which continue to constrain such investigations today).

 \bigskip

\noindent {\bf 2. Bohr's initial arguments.}\\

It should first be noted that the original EPR experiment involving position and momentum has some
significant differences from the more commonly discussed later version due to Bohm, the latter based on a spin- 1/2 singlet state. In the former case, measuring one observable involves a coupling with its complementary quantity, while that is not the case with the latter spin experiment. In the spin case, however, it can still be argued that the measurement conditions suitable for one spin observable are incompatible with measurements of a non-commuting spin observable.

With that in mind, let us attempt to distill Bohr's much-analyzed reply to EPR down to its essence. First,
consider his discussion of measurement of a single quantum system S's position or momentum using a
diaphragm D (screen with a single slit in it). The basic thought experiment can be described as follows:
\smallskip

1. Assume that S has an initial well-defined longitudinal momentum $p$ , with zero transverse component
(corresponding to the plane of the diaphragm), as it approaches the diaphragm D with slit.

2. Note that upon exiting D, S's state is one with greatly decreased transverse position uncertainty $\Delta q$ and
correspondingly increased transverse momentum uncertainty $\Delta p$.

3. Ask whether one could `foil' the uncertainty relation by taking into account any exchange of momentum
between S and D in order to reduce the uncertainty  $\Delta p$.

4. Assert that this is impossible because the exchange of momentum is `uncontrollable.'
\smallskip

Regarding (4), Dickson [5] notes that the characterization of the exchange of momentum as `uncontrollable'
is basically ``an article of faith'' on Bohr's part, and suggests that one should more conservatively call the
momentum exchange `unpredictable.'

What remains ill-defined in Bohr's account is whether the uncertainties and unpredictabilities in the
measurement processes are to be understood as genuine ontological indeterminacies or merely epistemic
ignorance of determinate values. This, I suggest, is a crucial equivocation in Bohr's treatment of the problem.
When dealing with objects that are decidedly quantum systems (such as the particle S going through the
slit), he seems to allow (at least implicitly, at this stage) these incompatible properties to be fundamentally
indeterminate. On the other hand, when dealing with macroscopic systems, he uses epistemic language,
referring to the relevant interactions and properties as `uncontrollable,' `inaccessible,' `unpredictable,' etc.
This is so even when he argues that under certain circumstances even a macroscopic object such as D should
be considered one of the quantum `objects of study.'  Such a circumstance would apply, for example, to his
proposal to delay the final measurement of D's displacement after passage of S and leave it as a matter of
`free choice' -- thus treating S and D as quantum-entangled in the EPR sense.

Of course, this equivocation concerning the nature of uncertainty (ontological vs. epistemic) serves to
evade the undesirable result that a macroscopic object like D could have a genuinely indeterminate position; if
one pursues that line avenue of inquiry, we are led immediately to the Schr{\"o}dinger Cat paradox (more on that
in the next section). One might argue that, even if taken as ontologically indeterminate, under the discussed
thought-experiment the indeterminacy of D's displacement would be so tiny as to be effectively microscopic
and therefore not observable. But one could, at least in principle, reversibly amplify the displacement of D to
macroscopic proportions, in which case D would be in many places at once. Bohr clearly does not accept this
idea; thus he must take position uncertainty pertaining to D as epistemic.

With regards to statement (3) above, Bohr notes that what makes step (2) a position measurement is
that D is anchored immovably to the lab frame, which establishes a spacetime frame of reference. Without
that spacetime frame, the notion of a position value would be ill-defined even in a classical sense. This
however is yet another form of equivocation on Bohr's part. As Dickson further points out, ``there is nothing
that, quantum mechanically, can really serve to define a reference frame, because reference frames are (by
definition!) well-defined both in position and momentum. Quantum theory tells us that there is no such thing,
but for the sake of making our notions of position and momentum meaningful, we voluntarily choose to
accept a given physical object (the apparatus, or whatever) to serve as a reference frame.'' ( [5] p.14, my
emphasis)

Here we encounter a form of the dilemma faced by Bohr and, I hope to persuade the reader, ultimately
not resolved by his notion of Complementarity. Our world of experience is clearly classical in that we can
legitimately consider our lab and macroscopic measuring instruments as inhabiting a well-defined inertial
frame. \textit{But these are the very phenomena that cry out for explanation in view of that fact that
the microscopic quantum objects upon which we experiment, according to the theory describing them, do not inhabit well-defined reference frames. }(This seemingly paradoxical situation actually is amenable to
resolution, which is the subject of Section 4.) Bohr deals with these apparently disparate realms
by equivocation concerning their physical nature, and that equivocation is aided by his use of qualitative
description rather than quantitative application of the relevant theoretical formalism. The next section aims to
remedy this reliance on qualitative description in order to more clearly identify the underlying weaknesses in
Bohr's account.

\bigskip

\noindent {\bf 3. Analysis of Bohr's second thought experiment}\\

Suppose we apply the quantum formalism to Bohr's thought experiment of the second case considered; i.e.,
the case in which D is allowed a transverse degree of freedom in order to have the possibility of measuring
the momentum of S. (This is termed experiment A-2 by Bai and Stachel.) This is to be done via momentum
conservation by measuring D's momentum before and after passage of S. But according to Bohr, this leaves
us with the `free choice' whether or not to measure position instead of momentum, by choosing whether or to
make a final position or momentum measurement of D. Thus Bohr seems to be describing the interaction
between S and D as a non-disturbing `measurement of the first kind,' sometimes termed a `pre-measurement'.
The initial state of S is $| p \rangle$, a state of well-defined momentum with zero transverse component,  and and D is in a ready state of well-defined position $|Q\rangle$. After their interaction, Bohr seems to assume that they can be represented by an entangled state $ | \Psi\rangle$ (much like the original EPR state, as noted by Bai and Stachel). As observed by EPR, such a state has an inherent basis ambiguity and can be written in terms of any orthogonal set of states. For reference, equations (7) and (8) of the original EPR paper are reproduced here:

	$$ \Psi(x_1, x_2) = \Sigma_n \psi_n(x_2) u_n(x_1)   \eqno(1)  $$ \

\noindent where $\Psi(x_1, x_2)$ is a two-particle wave function expressed in terms of eigenfunctions $u_n(x_1)$, of some observable A; and the coefficients $\psi_n(x_2)$ are viewed as amplitudes for the expansion in this basis.

On the other hand, as EPR note, the same two-particle state can be expressed in terms of a different set of eigenfunctions $v_n$ corresponding to a different observable B, with different expansion coefficients:

$$ \Psi(x_1, x_2) = \Sigma_n \phi_n(x_2) v_n(x_1)   \eqno(2)  $$ \

Let us define EPR's first observable A as applying to relevant aspects of the position of D. For convenience, take the eigenstates to be a discrete set of small transverse position ranges $| Q_i \rangle $ (one of which would act as a pointer to the localized wave packet $|q_k \rangle$ emerging from the slit). The second observable B will apply to the transverse momentum of D and its eigenstates will be a discrete set of small transverse momentum ranges $| P_j \rangle $ (which would act as a pointer to the transverse momentum state $|p_j\rangle $ of the emerging particle). The corresponding discrete states for S will be $| q_i \rangle $ and $| p_j \rangle $ respectively. So  $ | \Psi\rangle$ will look like:

$$ | \Psi\rangle \sim  \Sigma_i \alpha_i | q_i \rangle | Q_i \rangle  =  \Sigma_j \beta_j | p_j \rangle | P_j \rangle \eqno (3) $$

\noindent where $\alpha_i$ and $\beta_i$ are amplitudes. Thus, if we choose to measure the final position Q of D and find it within the range $ Q_k $, then the `entire experimental arrangement' allows us to attribute to S the corresponding state $| q_k \rangle$; or if we measure the final momentum P of D and find it within the range $ P_n $, then similarly in virtue of Bohr's `wholeness' criterion, we can attribute to S the corresponding state $| p_n \rangle$. 

Now, presumably the designated unitary evolution of the initially independent systems to the above
entangled state would have to be treated as a correlation arising via scattering of the particle from the edges
of the slit in D. To get some feel for the magnitudes involved, assume an incoming electron energy of roughly
 20 Mev, and a mass for D of as little as 1 gram (small but still macroscopic), in an elastic scattering process. 
 The maximum possible outgoing velocity for D would be negligible: of the order of $10^{-17}$ m/s. This is good
and bad news for Bohr. The good news is that such a microscopic effect might make it seem reasonable
to consider S and D as two entangled quantum systems (D being on the same footing as S as an `object of
study'). But the bad news is that D could not serve as a credible measuring instrument for the momentum of
S, and therefore we would not really have a `free choice' at this point to make that measurement given the
putative entangled system as described. In order to accomplish the latter, and still provide the ``free choice''
that Bohr asserts, D would need to be entangled with some sort of amplifying degree of freedom. But in that
case, we have a Schr{\"o}dinger's Cat situation: any indeterminacy in either D's position or momentum would be
visible at the macroscopic level, but it never is.

Thus, we can see that this is just the usual problem in which macroscopic objects, when assumed to be
described by quantum states entangled with quantum systems, become `infected' with any indeterminacy
pertaining to the quantum system. That is, it is the measurement problem. As Dickson notes: ``Presumably,
to consider the interaction between [the particle] and the apparatus a genuine measurement we must ignore
the subsequent entanglement between them and take the apparatus to be in a definite state of indication,
even if in fact it is not.'' ( [5] p. 28, preprint version.) This inconsistency problem is not addressed by the
notion of Complementarity. That is, it is fine to note that certain observables are incompatible and cannot
be simultaneously measured, and that it may be inappropriate to regard values of such sets of observables
as all well-defined under specified circumstances. But since such an observation does not resolve the above
consistency issue, it would appear to amount to little more than just restating the uncertainty principle.
``Complementarity'' is not enough.\\

\noindent {\bf 4. Bohr's epistemological and methodological assumptions as unnecessary restrictions on his
investigation}\\

At this point we consider some methodological and epistemological pronouncements by Bohr, which
represent the constraints under which his investigation took place, but which can in fact be questioned.
It should first be noted that the often-emphatic nature of Bohr's assertions may be understood as a legitimate
response to the need to question the classical, mechanistic, reductionist thinking prevailing at the time.
That is, Bohr was correct to emphasize that quantum theory represented a wholly new type of epistemological
and ontological challenge for the practice of physics as it had been traditionally understood; however, he
himself was also operating under certain preconceptions. Thus, although one can understand the categorical
and emphatic tone of some of his statements in the historical and philosophical context in which he was working,
one must also approach his assertions critically, not regarding them as the last word on the subject.

Consider now this rather lengthy categorical assertion:
\smallskip

{\small \textsf{``The essential lesson of the analysis of
measurements in quantum theory is thus the emphasis on the necessity, in the account of the phenomena,
of taking the whole experimental arrangement into consideration, in complete conformity with the fact that
all unambiguous interpretation of the quantum mechanical formalism involves the fixation of the external
conditions, defining the initial state of the atomic system concerned and the character of the possible
predictions as regards subsequent observable properties of that system. Any measurement in quantum theory
can in fact only refer either to a fixation of the initial state or to the test of such predictions, and it is first the
combination of measurements of both kinds which constitutes a well-defined phenomenon.''} }[11]\\

One can make the above assertion considerably less lengthy. Omitting some of the categorical and emphatic
aspects, the basic claims are found to be:\\

1. Measurement in quantum theory can only be physically defined by reference to a macroscopic
experimental arrangement.

2. A well-defined phenomenon, taken as defining `measurement,' requires an initial preparation and final
(macroscopic) observation.

3. There is no unambiguous interpretation of the quantum formalism as applied to any system without
reference to externally fixed conditions defining the initial and final states of that system, where `externally
fixed conditions' means macroscopic phenomena accessible to an observer.\\

In what follows, I critique these claims. A refutation of all three is presented in the final section, through a
counterexample: a formulation that unambiguously specifies how the determinacy inherent in measurement arises without necessary reference to macroscopic phenomena.

Firstly, while Bohr's insistence on the ``necessity... of taking the whole experimental arrangement into
consideration'' is well known, and is often taken as a benign statement of `quantum wholeness,' it is actually
a very strong (and, I will argue, unnecessary) prohibition on taking any degree of freedom as physically
specifiable independently of macroscopic phenomena. This prohibition is sharpened in claim 3 which effectively
asserts that one is not allowed to say that the quantum formalism, as applied to any subsystem of an `entire
experimental arrangement,' has an unambiguous physical referent, even if one cannot describe that referent in
``ordinary''-- meaning classical -- terms. Note that this is a stronger claim than merely saying ``an unmeasured subsystem does not have classically observable properties''; rather, it says that one should not try to understand the physical nature of any degrees of freedom that are \textit{correctly} assigned a quantum theoretical description. 

Overall, Bohr's quoted statement assumes that unambiguous physics only obtains in the context of a `measurement,' where that term is considered to be definable only in terms of a macroscopic
experimental arrangement leading to an `observation' or `phenomenon'. This use of the term `measurement'
is a conflation, ongoing in much of the literature, of two distinct ideas: (i) the intervention of an observer
whose intent is to gain determinate knowledge about something under study; and (ii) the existence of a fact
of the matter -- or determinate a value of some property -- whether or not anyone has intent to discover it (or
whether or not it results from a macroscopic `phenomenon'). The preceding two different notions of the
determinacy obtaining in measurement (but not necessarily confined to a knowledge-gathering measuring
operation) can be labeled as (i) epistemic and (ii) ontological, respectively. Bohr's pronouncement of course
denies (ii) by asserting that it is only through an in-principle macroscopic `phenomenon' that any physical
quantity is well-defined, and that the quantum formalism is not even interpretable outside that condition. But
this denial can and will be questioned.

Besides the above conflation, Bohr's insistence that one must take ``the whole experimental arrangement
into account'' does not remedy the consistency problem concerning S and D in their purported entanglement
that he describes in his reply to EPR. One supposedly has a ``free choice'' whether to measure the momentum
of D and thereby gain knowledge of the momentum of S on passing through the slit, or to measure the position
of D and thereby gain knowledge of the position of S. In this case S and D are in an entangled pure state and
D and S are described by improper mixed states. There is no basis from within the theory to say why, at the
time when the choice is supposedly available, any uncertainty pertaining to D should be of a different sort
than that pertaining to S. Yet clearly Bohr needs D's uncertainty to be epistemic rather than ontic in nature
to avoid a Schr{\"o}dinger's Cat situation; while on the other hand, since he views any attributes of a quantum
system such as S in need of (at least) irreversible amplification [10] in order to be considered determinate, the
uncertainty pertaining to S cannot be considered epistemic. However, the theoretical description provides no
justification for attributing different sorts of uncertainties to S and D.

Ultimately, Bohr's response to this conundrum is to deny reality to quantum objects, and to assert by fiat
that at some point in the (assumed as linear) evolution, a determinate world of experience occurs and classical
`reality' begins -- since we routinely see objects like D with determinate position and momentum. This is not an
explanation of classical emergence, but rather an equivocation concerning the application of quantum theory.
A crude analogy is that the unitary quantum evolution is like a car engine engaged via the clutch with the gear
shaft (which carries the entanglement of the relevant degrees of freedom); but at the point in which we find
ourselves empirically describing objects that are classically determinate (or, in which the dimensions of the
experiment are much larger than Planck's constant), we disengage the clutch. This is an \textit{ad hoc} move; there is
no consistent theoretical account for suspension of the unitary evolution. (It will not do to reply, in Bohrian fashion, that ``the lesson of quantum theory is that there can be no consistent theoretical account,'' since one is
provided in the final section.)

However, could we see this sort of move as justified by seeing it as form of pragmatism? I think the
answer is negative, and arguably does a disservice to pragmatism. Pragmatism primarily concerned itself
with reforming the concept of truth from an abstract and absolute notion to a concrete and functional one. It
is one thing to say that our criteria for truth must require that truth claims pass some test of functionality; it
is quite another to suspend the quantum formalism to force the theoretical description to correspond to our
empirical experience and/or to classical mechanics at a specified limit, even though it apparently does not.
That is the essence of equivocation, and pragmatism was not equivocal.

It is worth mentioning in this context that Bub [8] has given an interesting formal account of Bohr's ``Complementarity.''
Bub has shown that the Hilbert space structure of quantum states allows for a generalization of
the ``Bohmian'' theory in which the position of a quantum system is taken as an always-determinate ``beable''
(Bell's term, [9]). It turns out that one can always choose one particular observable as having preferred status,
such that its eigenvalues attain ``beable'' status; and any other observable commuting with that preferred
observable will have determinate values as well. Meanwhile, properties corresponding to observables not
commuting with the preferred observable have indeterminate status (there are no yes/no answers to questions
about those properties, where the questions are represented by projection operators on the Hilbert Space).
According to Bub's observation, Complementarity consists of conferring ``preferred'' status on the observable
selected as being determinate by the `entire experimental arrangement,' such that its eigenvalues become
``beables.''

Does this allow Bohr to escape from the above inconsistency problem? I believe the answer is ``no''. Recall
that Bohr says we have a free choice whether to measure position or momentum of the diaphragm in the
case in which S and D are assumed to be entangled; he asserts that D is to be viewed as a quantum system
at this stage of the experiment. Clearly the availability of this ``free choice'' means that we have not yet
completed the ``entire experimental arrangement'' that would bring about a preferred observable according
to Bub's formulation. But this means that (at this stage of the experiment, prior to the choice), there is
no fact of the matter about either D's position or its momentum, since neither is a preferred observable.
Thus invoking a preferred observable-based beable does not rescue Bohr from the inconsistency, since his
``entire experimental arrangement'' criterion for the preferred observable implies the undesirable conclusion
that at certain preliminary stages of an experiment, a macroscopic object has no determinate physical
property. It should be kept in mind that the tiny displacement of D does not help here: according to Bohr's
assumptions, in principle one could reversibly entangle another degree of freedom with D that would amplify
the tiny displacement to macroscopic proportions and yet still be described, according to Bohr, as a quantum
system (since there has been no ``irreversible amplification'' such as a change in the chemical properties of
photographic plate emulsion that Bohr takes as heralding a ``measurement'').

As noted in the Introduction, Bohr's views evolved over time. For example, as Stachel points out, ``Bohr's
later approach places primary emphasis on four-dimensional processes; from this point of view, a `state' is
just a particular spatial cross-section of a process, of secondary importance: all such cross-sections are equally
valid, and any such sequence of states merely represents a different `perspective' on the same process.'' ([12], p. 1, preprint version.) It should however be noted that such an approach -- dissolving the measurement problem by noting that some outcome always in fact obtains at the phenomenal, classical, spacetime level -- amounts
to an epistemic interpretation of the quantum state. That is, the quantum state and its unitary evolution are
taken as describing only our limited perspective on a process that is assumed to be complete as an element of
a classically determinate block world. In this approach, the classical world of phenomenal experience does not emerge from the quantum level. It is taken as ontologically given and primary, with quantum theory relegated to a partial and
perspectival description of that classical reality.\footnote{Stachel (private communication) gives another argument for denying ontological reality to the quantum state. This consists in the observation that a time-symmetric approach to the Born Rule will attribute a different state to the same
system depending on whether it is considered a pre-selected or post-selected. In terms of the Aharonov-Bergmann- Lebowitz rule [13], this is seen in the fact that the ABL rule gives a probability of unity for an intermediate measurement
of either the pre- or post-selected state. But what this implies for interpretation of the quantum state depends crucially
on one's presumed ontology. If one presumes that there is a block world (i.e. no ontological difference between past,
present, and future), then the foregoing results simply restate that ontology, since in a block world each system is both
prepared and fated at any intermediate time during its lifetime. On the other hand, in a growing universe ontology with
indeterminate future, the foregoing results do not indicate any inconsistency for an ontological quantum state. The
prepared state can be understood as describing the system prior to its detection, while the attribution of the post-selected
state is only applicable a posteriori. (This is essentially the case for the interpretation to be discussed in the final section.)}

In addition to the pronouncement which opened this section, Bohr made many other emphatic, categorical
statements concerning the interpretation of quantum theory that are nevertheless subject to challenge as being
based on (a) unacknowledged metaphysical and conceptual premises, or (b) even on an ill-defined ontology.
An example of (a) is the following:\\

{\small \textsf{``It must not be forgotten that only the classical ideas of material particles and electromagnetic waves have
a field of unambiguous application, whereas the concepts of photons and electron waves have not. Their
applicability is essentially limited to cases in which, on account of the existence of the quantum of action, it is
not possible to consider the phenomena observed as independent of the apparatus utilized for their observation.
I would like to mention, as an example, the most conspicuous application of Maxwell's ideas, namely, the
electromagnetic waves in wireless transmission. It is a purely formal matter to say that these waves consist
of photons, since the conditions under which we control the emission and the reception of the radio waves
preclude the possibility of determining the number of photons they should contain. In such a case we may
say that all trace of the photon idea, which is essentially one of enumeration of elementary processes, has
completely disappeared.'' }} [14], 691-92.\\

The phrase ``electromagnetic waves in wireless transmission'' means the classical electromagnetic field. Such a field
is instantiated by the quantum coherent state, which is a superposition of photon number. To obtain
a detectable classical field, one needs a very large average photon number.\footnote{Sakurai notes that ``The classical
limit of the quantum theory of radiation is achieved when the number of photons becomes so large that
the occupation number may as well be regarded as a continuous variable.'' ([15], p. 36)} Note that Bohr has slid from the fact that
the coherent state is a quantum superposition of photon number to the conclusion that ``the photon idea has
disappeared''. But it has not: the coherent state can be understood as a well-defined physical quantity, whether
or not that it is visualizable ``in the ordinary (classical) sense''. [14], p.21. He thus simply disallows an ontology
in which there could be a physically real state of the field, involving an indeterminate number of photons,
that is not visualizable in a classical way. But, as Ernest McMullin has noted, ``[I]maginability must not be
made the test for ontology. The realist claim is that the scientist is discovering the structures of the world; it
is not required in addition that these structures be imaginable in the categories of the macroworld.'' [16],14.
In contrast, Bohr routinely insisted on the latter condition as a basic methodological requirement for doing physics.
Moreover, that condition is precisely his criterion for what is to be regarded as physically real: according to
Bohr, what is not ``visualizable in the usual (classical) way'' is deemed ``abstract'' and even ``undefined,'' as
we will see further below in considering an example of (b) (an ill-defined ontology).

Thus, Bohr's assertion peremptorily rules out \textit{even the possibility} of an unambiguous physical referent for the key theoretical objects of quantum theory -- discrete quanta and de Broglie waves. Yet it is dependent on the implicit and unnecessary assumption that all real physical processes must be classically visualizable spacetime processes, and on the accompanying assumption that quantum discreteness can only mean spacetime localizability as a `corpuscle'.

The statement was made in the context of Bohr's inability to reconcile the idea of a wavelike frequency
with the presumed corpuscular idea of a `photon', and the inverse problem of specifying within spacetime any
wavelike (extended) nature of a `material particle' such as an electron. But Bohr's negative conclusion is not
forced on us: a quantum of electromagnetic radiation or `photon', as the singular entity heralding a quantum
discontinuity, need not be considered as a spatially localized object. The quantized, indivisible aspect of
the photon can be reinterpreted as a component of the process of emergence of spacetime events and their
discrete connections, the photon being the latter. Thus the discrete photon can be understood as emerging under certain suitable
physical conditions, and the coherent state discussed above can be understood as a pre-emergent form of
the underlying field. Meanwhile, the wavelike character of the photon and other material quanta (i.e. the de
Broglie oscillation) can be retained on a sub-empirical, pre-spacetime level.

Such an approach, in which quantum processes are precursors to the emergence of localized spacetime
events and their connections, is briefly reviewed in the final section. (It should also be noted that the present
author is not the only one currently exploring spacetime emergence; cf. Sorkin [17], Oriti [18].) 
Thus, with a suitable relaxing of conceptual barriers and unnecessary metaphysical presumptions, one can indeed gain an
unambiguous application for the basic physical concepts of quantum theory, contrary to Bohm's categorical
negative claim.\

An example of (b), a statement from Bohr exhibiting an ill-defined ontology is: \\
{\small \textsf{``Isolated material particles
are abstractions, their properties being definable and observable only through their interaction with other
systems.''}} [19]\\

This statement is problematic in several ways. First, many abstractions are perfectly well defined (such as
mathematical concepts); so lack of definition has nothing to do with whether or not something is abstract. But
more importantly, how does a non-physical, allegedly undefined abstraction undergo physical interactions?
And if the interactions are not physical, how does a process that \textit{could} be deemed concrete and physical come out of any of that?
This is essentially the same ``remove the clutch'' inconsistency encountered above, where Bohr describes
the initial degrees of freedom (S and D) by an entangled state and its unitary evolution, but then assumes that
something real and determinate (somehow) occurs so that at least one of the same degrees of freedom (D) is
no longer described by the entangled quantum state and its unitary evolution. There is a gap between the
allegedly `abstract and ill-defined' and the allegedly `non-abstract and well-defined' that is not bridged by
any amount of `amplification.' This problem can be seen as the same type of metaphysical inconsistency
facing Cartesian mind-matter dualism in that one has two fundamentally different substances that have no
way to `interact.'

In an epistemic approach to the quantum state, Bohr could finesse the inconsistencies described above
by saying that we can suspend unitary evolution when it is no longer useful because we now have access to
information that we lacked previously. Thus, neither the quantum state nor its unitary evolution ever directly
described objects that physically existed. All that exists is the phenomenal, classical level of experience. But
again, this leads Bohr to his ultimately antirealist view of quantum entities; i.e., to his utterance that ``There
is no quantum world. There is only an abstract quantum mechanical description.'' If there is no quantum
world, then we need not give any account of classical emergence from such a world, since all that exists is the
classical world of experience.

Bohr can thus retain a kind of consistency, but only (at least it seems to this author) at a rather high cost.
Bohr spent the bulk of his career developing a detailed and revolutionary theory of the hydrogen atom in terms
of its applicable quantum states. In order to retain consistency in the face of reconciling quantum mechanics
with the classical realm of experience under the assumption of unitary-only evolution, Bohr ultimately felt
forced to deny that hydrogen atoms could count as real physical referents for the very quantum states that he helped to formulate for them.
 Perhaps this turn to antirealism about the constructs of his pioneering theory was not really
necessary. We consider an alternative in the next and final section.

Before turning to that alternative, it should be noted that Bohr's formulation legitimately takes measurement
and determinacy as contextual; but it goes further than that by presuming that the contextuality is necessarily
always a macroscopic one, dependent on a ``phenomenon.'' The latter term essentially means ``appearance,''
and thus is an intrinsically observer-dependent notion (since any appearance is always relative to a perceiving
subject or subjects). This is a symptom of the fact that Bohr is unable to say why only one outcome occurs
if one applies linear evolution to a quantum system and all its correlates; that of course always leads to a
Schr{\"o}dinger's Cat situation. So Bohr instead assumes that one must start with the observer's experience,
where only one outcome is perceived; then one at least apparently has a well-defined physical situation.

But it is not in fact the case that this is the only way to obtain a well-defined physical quantity, and therefore
it is not necessary to appeal to macroscopic `phenomena' as an ostensibly necessary starting point. The
fundamental unnecessary constraint on Bohr's thinking is the presumption that the condition giving rise to a
determinate value of a quantum mechanical operator cannot be defined from within the quantum formalism
alone. But in fact it can.

\bigskip

\noindent {\bf 4. Beyond Complementarity.}\\

The above-discussed apparent discrepancy between theory and observation, to which Bohr's Complementarity
and its attendant antirealism about quantum objects is sometimes taken as a perplexing but inescapable
response, is not a necessary one. The problem arises from demanding that all interactions between physical
degrees of freedom are unitary ones. This is the key assumption that leads to the measurement problem and
the ``shifty split'' between the quantum and classical realms, expressed in the \textit{ad hoc} suspension of the unitary
evolution and quantum-entangled state when it obviously no longer correctly describes the situation at hand.
If nature in fact involves real non-unitary processes of a well-defined sort -- including the circumstances that
give rise to them -- then the chain of unitary correlations is broken, and real physical collapse occurs, resulting in determinacy. Thus,
the present author suggests that what Bohr needs to avoid the dilemma of theoretical inconsistency on the one
hand, and antirealism about quanta on the other, is genuine, non-unitary physical collapse.

What is also needed is an expansion of our metaphysical notions concerning what qualifies as `physically real' -- specifically, the
acknowledgment that there may be real entities, referred to by the theoretical constructs such as quantum states, that are not be confined to 3+1 spacetime. 
Thus the present proposal differs with the ``primitive ontology'' (PO) approach discussed by Allori [20]: the starting point for the PO
is the assumption that any fundamental ontology referred to by a theoretical construct must be an element of the spacetime manifold.
This restriction under PO of the ``primitive variables'' to 3+1 spacetime is prompted by the following consideration: \\

{\small \textsf{``Roughly, the three-dimensionality of the primitive variables allows for a direct contact between the variables in the theory and the objects in the world we want them to describe. In fact, a PO represented by an object in a space of dimension d, different than 3, would imply that matter lives in a d-dimensional space. Thus, our fundamental physical theory would have to be able to provide an additional explanation of why we think we live in three-dimensional world while we actually do not.'' }}[20]\\

The proposed solution to this challenge is that quantum states refer to sub-empirical, pre-spacetime entities that 
can constitute precursors to observable spacetime events.\footnote{The question of why the observable spacetime manifold
is 3+1 dimensions is a deep one with many different proposed answers. One relevant fact is that photons, which create observability, have 4 polarization directions.
But for our purposes, it is sufficient to note that observable processes are always spacetime phenomena, while intrinsically unobservable quantum
processes need not be required to inhabit the same manifold as the observable ones, as long as an account can be given of the transition
from one manifold to the other. This is indicated in Kastner 2012 [ 21] and later in this section; the transition is precisely the collapse process.}
 That is, the ontology has distinct levels: (i) actuality
(observable, element of the spacetime manifold) versus (ii) physical possibility (still real but unobservable, pre-spatiotemporal).
Level (ii) is essentially the Heisenbergian ``potentiae'' [22]. Such an ontology, to be described in more detail below, is consistent with the reasonable view that real entities should be capable of leading to observable results, even if they themselves are not observable. In fact the latter view is attributed to Bohr by Bai and Stachel, who say : ``[Bohr's] (and Einstein's) view is that what exists must
be measurable, or more accurately, must have measurable consequences.''[1] However, despite this apparent  initial openness to allowing physical existence to non-classical, unobservable entities, Bohr steadily evolved
toward a form of antirealism that denied reality to objects not in-principle capable of a classical description,
i.e. ``which cannot be visualized in the ordinary sense'', as his above-quoted assertions clearly demonstrate.

Returning now to the need for real collapse: there are `spontaneous collapse' models out there, the best
known being that of Ghirardi, Rimini, and Weber [23]; but these involve changing the Schr{\"o}dinger equation
(by adding non-linear terms designed to bring about dynamical collapse). The model that does not modify the
basic quantum evolution (although it incorporates an additional step resulting in collapse) 
is based on the direct-action theory of fields, called the Transactional Interpretation
(TI) [21, 24]. TI defines the usual retarded solution to the Schr{\"o}dinger equation as an `offer wave' (OW). But it
also includes an additional process beyond the unitary evolution of the offer wave, namely an
 advanced response from absorbers. The advanced response, called a `confirmation wave' (CW), is a
solution to the complex conjugate Schr{\"o}dinger Equation. This response is what precipitates collapse by
breaking the linearity of the evolution of the quantum state (OW).

In general, one OW will elicit responses from many
absorbers, where each such absorber receives only a component of the original OW. (A typical example of
this is an interferometer experiment in which a beam splitter directs OW components to different detectors.)
The process of CW response to OW components corresponds to the von Neuman `Process 1' measurement
 transition from a pure state (the OW) to a mixed state (weighted projection operators corresponding to the
different OW components and their respective CW responses). As discussed in [21], \S 3.2.3, this mixed state represents a set of incipient
transactions, only one of which can be actualized. It is proposed that the ``collapse'' to one outcome among
the many (now in a well-defined basis due to the inclusion of absorber response) occurs through an analog of
symmetry breaking, which is ubiquitous in physics (cf. [25]) The actualization of the transaction constitutes a
transfer of measurable conserved quantities (energy, momentum, spin, etc.) from the emitter to the `winning'
absorber. In the transactional picture, a photon is just this transfer of in-principle detectable electromagnetic
energy, momentum, and angular momentum; and it is a discrete quantity where the energy 
$E = h\nu$.  Thus, there
is a real physical, nonunitary collapse in this model. There is also a clear physical referent for the ``photon''
concept independently of whether any macroscopic, observable ``phenomenon'' (involving an observer) results
from it.

TI has been extended by this author to the relativistic domain, together with an ontological reinterpretation
of the OW and CW as pre-spatiotemporal physical possibilities (reminiscent of Heisenbergian `potentiae' as noted above).
This version is called the `Possibilist Transactional Interpretation' (PTI) [21]. 
In this picture, the collapse is not a spacetime process (which is already known to be problematic [26]); rather, it is a discontinuous process by which spacetime events (actualities) emerge from  a quantum level of potentiality.
The current paper focuses on Bohr's
views, and will not present a detailed case for TI or PTI (that has been presented in [21], and also in [27]). The
point is just to note that Bohr's conclusions are not inevitable, since they are based on certain methodological and metaphysical assumptions and constraints that need not be accepted; and that they do contain gaps and equivocations, which can in principle be remedied in an appropriate non-unitary collapse model of measurement. 

However, in view of Bohr's rejection of the quantum coherent state as a purely ``formal'' construct in
which the ``idea of the photon is lost,'' it should be pointed out that [21], Chapter 6 discusses the physical
relationship between the coherent state and the classical electromagnetic field that emerges from it through
sustained actualized transactions. In this context, the term `photon' can also refer to the offer wave capable
of transferring one quantum of electromagnetic energy, and a coherent state is just an offer wave that is
capable of transferring a varying number $n$ of detectable photons where $n$ is characterized by a well-defined
probability. It is the fact that the coherent state is an eigenstate of the field destruction operator that allows it to
function in this way; the repeated absorption of photon(s) from the field does not change the field state, which
is what allows a detectable classical field to be sustained. So the photon as a physical entity remains quite
meaningful -- even crucial -- in the quantum coherent state. A detailed account of the well-developed theory
of coherent states, including experimental verification of the theoretical predictions for photon detections,
is found in [28]. To say that the ``photon idea disappears'' just because there is an indeterminate number of
photons in the pre-detection field is at variance with both theory and experiment on coherent states.

Another aspect of PTI should be mentioned here: recall the point made in Section 2 concerning the
inconsistency of defining an inertial frame of reference at the quantum level. This problem is remedied under
PTI by taking spacetime as an emergent structure, supervenient on actualized transactions between quantum level
emitters and absorbers. In order to describe this emergence, PTI takes literally the idea that energy
and momentum are the generators of temporal and spatial displacements, respectively. Thus an actualized
transaction resulting in the transfer from emitter x to absorber y of a quantum with energy E and momentum
p defines a spacetime displacement ($(y^{\mu} - x^{\mu}$) that is characterized by an invariant interval. (In the rest frame
of a transferred material quantum, $p = 0$ and there is zero spatial displacement; this defines the temporal
axis for the particle.) 

In addition, macroscopic (classical) objects are distinguished from quantum systems in a well-defined (although
inherently probabilistic way): they are overwhelmingly likely to bring about collapse, since they are huge collections of potential
emitters and/or absorbers ([29], \S 5 and for a non-technical presentation see [30], pp 96-106). This makes it virtually impossible to coherently entangle an object like D with
a quantum system S such that unitary evolution is preserved. This is a form of decoherence, but one based
on a physically irreversible process. As such it avoids the circularity problem of the traditional decoherence
program (cf. [31].) Moreover, in this picture, irreversibility arises naturally as a previously unsuspected law of nature; thus the second 
law of thermodynamics is explained without having to assume special low-entropy conditions. For example, thermal 
interactions are irreversible transactions, thus legitimizing Boltzmann's assumption of ``molecular chaos'' in his
derivation of his H-theorem.

Since a macroscopic object is a nexus of frequent and persistent transactions giving rise
to well-defined spacetime intervals, macroscopic objects can be described by simultaneous spacetime $(x,t)$
and dynamical $(E,P)$ descriptions, and as such are clearly distinguished from quantum systems described by
quantum states, which are elements of an underlying substratum. Thus we have classical phenomena in PTI
as well; they are simply a naturally emergent result rather than a necessary starting point in interpreting the
theory.

Concerning the matter of contextuality: Bohr was of course correct that one cannot simultaneously define
incompatible quantities when dealing with quantum systems. In terms of PTI, that is because determinate
physical quantities only obtain as a result of actualized transactions. The latter occur by way of specific
interactions between an OW and its responding CW. Confirmations define the basis for the measurement,
by setting up the applicable mixed state (for example, two weighted projectors corresponding to each of
two detectors in an interferometer experiment). Only the projectors in that mixed state are eligible for
spacetime existence (i.e. as transfers of detectable energy, momentum, etc.); so quantities corresponding
to noncommuting observables are simply not in play at that point. The CW thus constitute the physically
well-defined ``contextuality'' that Bohr felt forced to define only with appeal to final, external observations -- 
``phenomena''.

To emphasize the fact that such contextuality has nothing to do with macroscopic ``phenomena,'' an
example of a well-defined physical quantity under PTI is the energy/momentum of a photon emitted from
an excited state atom and absorbed by a ground state atom, regardless of whether that single photon is ever
amplified to the level at which it could in principle be perceived by a scientist in a laboratory. All the objects
involved are quantum systems, all described by quantum mechanics, and Planck's constant plays a crucial
role in the interaction. Yet there is an unambiguous interpretation of the quantum formalism, applying to
the degrees of freedom described by the formalism. No appeal to ``the entire experimental arrangement'' or
necessarily observable ``phenomenon'' is required for this interpretation. The context consists of any forces
acting on the photon offer wave (i.e., the applicable Hamiltonian) and the set of advanced absorber responses
to the photon offer (the latter being described by the usual forward-propagating quantum state). The context
is entirely physical. The transactional process, which heralds the advent of classicality (because it confers
determinate properties on the degrees of freedom involved) occurs at a microscopic level, independently of
whether any particular scientist is able to identify any macroscopic phenomenon arising from it.\\

\noindent {\bf 5. Conclusion.}\\

Complementary cannot help us to explain measurement or the nature of physical reality in a consistent
fashion unless we can explain why the quantum formalism applies correctly to quantum degrees of freedom
(such as the ``quantum particle'' S in Bohr's thought experiments with S and D) but not to macroscopic
objects; that is, why the ontic uncertainty of quantum objects does not ``infect'' macroscopic objects such as
Bohr's diaphragm D, and why we can view the latter's uncertainty as being epistemic. If we include absorber
response, we have a way forward to make this distinction in physical terms. Bohr was unable to do this
through Complementarity alone, and he lapsed into instrumentalist and anti-realist utterances as a result.

Recall Bohr's famous statement that ``It is wrong to think that the task of physics is to find out how nature is. Physics concerns what we can say about nature.''\footnote{These peremptory sentences followed Bohr's antirealist statement ``There is no quantum world. There is only an abstract quantum mechanical description.''}
Clearly, there is an implicit assumption here: ``physics cannot say how nature is.''  But in fact, quantum theory certainly can be telling us ``how nature is.'' Why should we presume that nature has to be determinate and classical at all levels, just because we cannot visualize it ``in the ordinary way''? 

Elsewhere in this volume, George Jaroszkiewicz [32] notes that the reductionistic assumptions behind classical physics need to be re-examined, and
 that physics is an empirical science. I certainly agree with both points. However, the fact that physical theory begins by engaging with empirical
 phenomena, and must be rigorously tested by experiment, does not negate the longstanding tradition in physics of theoretical description in terms of
unobservables. Boltzmann's atomic hypothesis is a prominent example.\footnote{Faraday's ``lines of force'' is another.} 
 It is well known that the idea of unobservable atoms was highly controversial,
and that Ernst Mach strongly objected to it on the basis that physics is an empirical science. Yet the atomic hypothesis was clearly the fruitful path,
and it is reasonable to take that theoretical success as evidence for the existence of atoms, especially now that we can (indirectly) image atoms. 

Similarly, it is reasonable to take the success of quantum
theory as evidence for the existence of additional structure in nature that gives rise to the kinds of phenomena predicted by the theory, even if it
is difficult (or even impossible) to visualize this structure ``in the ordinary (classical) way.'' This is ``inference to the best explanation'' for the 
empirical success of a theory. The new challenge from quantum theory is that such referents cannot be classical (i.e., not Einsteinian `elements of reality').
 But that in itself does not mean there can be no physical referent for the theory. 
  In contrast, an instrumentalist, observer-dependent interpretation of quantum theory can provide no explanation for the success of the theory
 in predicting (at the statistical level) our observations. It essentially says that we have a very good instruction manual for predicting the experiences of an observer, 
 but there is nothing in the world corresponding to the manual, and/or it is wrong to think there should be a reason or explanation for its predictive power. 
 Such an attitude would appear to be based
 on the assumption that if the explanation is not classical in nature (i.e. not in terms of determinate spacetime objects), there can be no explanation.
  But why should we demand that the explanation behind the success of quantum theory be classical? That expectation, I suggest, is what
  needs to be given up.\footnote{Of course, the disagreement between instrumentalists and realists can also be understood as a disagreement about the nature
  of scientific inquiry and explanation. Mach argued for a limited descriptive role for physical theory, and considered matters of ontology as strictly outside
  the domain of physics. However, such a methodological limitation on the discipline of physics does not \textit{preclude} reasonable ontological inferences 
  based on the success of physical theory, whether or not one considers such inferences as within the proper purview of physics. 
  And such ontological inferences may even prove fruitful in constructing new theories or in resolving anomalies or other remaining challenges in physics.
  This situation illustrates the ongoing fundamental dependence of physics on philosophy.}  
 
Finally, the proposed PTI picture of an intrinsically unobservable, pre-spacetime quantum substratum giving
rise to an empirical, classically determinate realm of experience may seem startling, even farfetched. But it
does provide a clear physical referent for the quantum formalism (at least in a structural sense, [33, 34]), and a well-defined basis for the emergence
of classical determinacy -- describable by classical physics -- from that formalism. In that regard, I
have noted elsewhere ( [21], Chapter 7) that the PTI ontology provides a natural correspondence for Kantian
``noumenon'' as describing the quantum level and ``phenomenon'' as describing the classical level. Here it is
advisable to recall again McMullin's observation that the structures of the microworld are not required to be
``imaginable in the categories of the macroworld.'' And as Bohr himself commented in a remark to Pauli, it
might just be ``crazy enough to be true.''\footnote{Bohr's famous remark concerning a theory by Pauli, as quoted in [35].}

\bigskip

{\bf Acknowledgments}\
I would like to thank John Stachel, Miroljub Dugi\' c, Jasmina Jekni\' c- Dugi\' c, Christian DeRonde, Bernice Kastner, and George Jarozskiewicz for valuable discussions.

\newpage

{\bf References}\bigskip

[1] Bai D, Stachel J. Bohr's Diaphragms. Forthcoming in this Volume, 2016.

[2] Beller M, Fine A. Bohr's Response to EPR. In: Faye J., Folse H., eds., \textit{Niels Bohr and Contemporary Philosophy}.
1-31. Dordrecht: Kluwer Academic Publishers, 1994.

[3] Bohr N. Can Quantum-Mechanical Description of Reality Be Considered Complete? \textit{Physical Review 48},
696-702, 1935.

[4] Einstein A., Podolsky B, Rosen N. Can Quantum-Mechanical Description of Reality Be Considered Complete?
\textit{Physical Review 47,} 777, 1935.

[5] Dickson M. Bohr on Bell: A proposed reading of Bohr and its implications for Bell's Theorem. In: \textit{Proceedings
of the NATO Advanced Research Workshop on Modality, Probability and Bell's Theorem}, Butterfield J, Placek T
(editors), Amsterdam: IOS Press, 2002.

[6] Letter to Paul Dirac (1928) In: \textit{Niels Bohr: Collected Works}, Kalckar J (editor), Vol. 8, pp. 44-46. North-Holland.

[7] Petersen A. The Philosophy of Niels Bohr. In: Bulletin of the Atomic Scientists, vol. 19, No. 7, 1963.

[8] Bub J. \textit{Interpreting the Quantum World}. Cambridge: Cambridge University Press, 1997.

[9] Bell J S. A Theory of Local Beables. CERN Ref. 2052, Presented at the 6th GIFT Seminar, 1975.

[10] Bohr N. \textit{Atomic Physics and Human Knowledge.} New York: John Wiley Sons, 1958, p. 88 (originally published
1955).

[11] Bohr N. The Causality Problem in Atomic Physics. In \textit{International Institute of Intellectual Cooperation}, 1939, pp.
11-30.

[12] Stachel J. It Ain't Necessarily So: Interpretations and Misinterpretations of Quantum Theory. Forthcoming in this Volume. 2016.

[13] Aharonov Y, Bergmann P, Lebowitz J. Time symmetry in the quantum process of measurement. Physical Review B
134,1410?1416, 1964.

[14] Bohr N., in Kalckar J, Ed. \textit{Collected Writings. Vol 6. Foundations of Quantum Physics I. }Amsterdam, North-
Holland/Elsevier, 1985.

[15] Sakurai, J J. \textit{Advanced Quantum Mechanics}. Reading: Addison-Wesley, 1973.

[16] McMullin E. A Case for Scientific Realism. In Leplin J (ed) \textit{Scientific Realism.} Berkeley: UCLA Press, 1984, 8-40.

[17] Sorkin, R. Relativity does not imply that the future already exists: a counterexample.
Vesselin Petkov (editor), \textit{Relativity and the Dimensionality of the World}. (Springer 2007 (arXiv:gr-qc/0703098)

[18] D. Oriti, ``A quantum field theory picture of simplicial geometry and the emergence of spacetime,'' in the Proceedings
of the DICE 2006 workshop, Piombino, Italy, Journal of Physics: Conference series.

[19] Bohr N. \textit{Atomic Theory and the Description of Nature}. Cambridge: Cambridge University Press, 1934.

[20] Allori, V. Primitive Ontology and the Classical World. This Volume, 2016.

[21] Kastner R E. \textit{The Transactional Interpretation of Quantum Mechanics: The Reality of Possibility.} Cambridge:
Cambridge University Press, 2012.

[22] Heisenberg, W. Physics and Philosophy.  New York: Harper Collins, 1962,  p. 41.

[23] Ghirardi GC, Rimini A, Weber T. Unified dynamics for microscopic and macroscopic systems. \textit{Physical Review
D34}, 470, 1986.

[24] Cramer J G.  The Transactional Interpretation of Quantum Mechanics. \textit{Reviews of Modern Physics 58}, 647-688,
1986.

[25] Brading, Katherine and Castellani, Elena, ``Symmetry and Symmetry Breaking'', The Stanford
Encyclopedia of Philosophy (Spring 2013 Edition), Edward N. Zalta (ed.), URL =
¡http://plato.stanford.edu/archives/spr2013/
\ entries/symmetry-breaking.

[26] Aharonov, Y. and Albert, D. Can we make sense of the measurement process in relativistic quantum mechanics?  \textit{Phys Rev D 24}, 359-70, 1981.

[27] Kastner R E. Haag's Theorem as a Reason to Reconsider Direct-Action Theories. \textit{International Journal of Quantum
Foundations 1}, No. 2, pp. 56-64 (2015).

[28] Breitenbach G, Schiller S, and Mlynek J. Measurement of the quantum states of squeezed light. \textit{Nature 387}, 471
(1997).

[29] Kastner R E. The Possibilist Transactional Interpretation and Relativity. \textit{Foundations of Physics 42}, pp. 1094-1113
(2012).

[30] Kastner R. E. \textit{Understanding Our Unseen Reality: Solving Quantum Riddles}. London: Imperial College Press. (2015).

[31] Kastner R E. Einselection of Pointer Observables: The New H-Theoreom? \textit{Studies in History and Philosophy of
Modern Physics 48}, pp. 56-58 (2014).

[32] Jaroszkiewicz G. Principles of Empirical Science and the Interpretation of Quantum Mechanics. Forthcoming in This Volume, 2016.

[33] Worrall, J. ``Structural realism: The best of both worlds?'' \textit{Dialectica 43}, 99 --124. Reprinted in D. Papineau (ed.), \textit{The Philosophy of Science}, Oxford: Oxford University Press, pp. 139 -- 165, 1989.

[34] Worrall, J. ``Miracles and Models: Why Reports of the Death of Structural Realism May Be Exaggerated''. \textit{Royal Institute of Philosophy Supplements 82}, (61): 125 -- 154, 2007.

[35] Scoular, S. \textit{First Philosophy: The Theory of Everything}. Universal Publishers, 2007.

\end{document}